\documentclass[aps,epsf]{revtex4}   

\def\bra{\langle}
\def\ket{\rangle}

\begin{document}

\title{  Nuclear Incompressibility in Asymmetric Systems
at Finite Temperature and Entropy}

\author{A.Z. Mekjian$^{a)}$, S.J. Lee$^{b)}$ and L. Zamick$^{a)}$}
                    
\address{$^{a)}$Department of Physics and Astronomy, Rutgers University
                             Piscataway, N.J. 08854}
\address{$^{b)}$Department of Physics and Institute of Natural Sciences,
                    Kyung Hee University, Suwon, KyungGiDo, Korea}



\begin{abstract}

 The nuclear incompressibility $\kappa$ is investigated
in asymmetric systems in a mean field model. 
The calculations are done at zero and finite temperatures
and include  surface, Coulomb and symmetry energy terms
for several equations of state.
Also considered is the behavior of the incompressibility at constant
entropy $\kappa_Q$ which is shown to have a very different behavior than the
isothermal $\kappa$. Namely, $\kappa_Q$ decreases with increasing entropy
while the isothermal $\kappa$ increases with increasing $T$.
A duality is found between the adiabatic $\kappa_Q$ and the $T=0$
isothermal $\kappa$. Analytic and also simple 
approximate expressions for $\kappa$ are given.

\end{abstract}

\pacs{ 
PACS no.: 21.65.+f, 21.30.Fe, 64.10.+h
 }

\maketitle
     

\section{Introduction}

Heavy ion collisions produce hot and dense hadronic matter \cite{dasgup,daniel}.
Such matter is also encountered in nuclear astrophysics and in
supernovae explosions.
Future rare isotope accelerators will explore nuclei at the limit of 
isospin asymmetry.
In neutron stars one also encounters system with a large neutron to
proton ratio.
Thermodynamic quantities play an important role in characterizing
properties of such systems.
One such quantity is the nuclear incompressibility.
While the nuclear incompressibility at zero temperature has been studied
for an extended period \cite{daniel,zamick,blaizot,pears},
it is only relatively recently that its temperature dependence
has been of concern \cite{muller,prlsub}.
In this paper we will focus on the temperature dependence of the
incompressibility in asymmetric systems, including the role of
surface, symmetry and Coulomb terms.
We will also study the adiabatic incompressibility which at $T=0$
would be the same as the isothermal incompressibility,
but differs from it at finite $T$.
The isothermal incompressibility will increase with $T$
while the adiabatic incompressibility will decrease with 
entropy \cite{prlsub}.

As a baseline, we will begin with
a mean field discussion of its behavior with $T$ to see 
the contributions to incompressibility of effective mass, excluded volume,
symmetry energy, surface energy and Coulomb energy.
In Sect.II, we will summarize a general description of incompressiblity
and results of Ref.\cite{prlsub} which considered the temperature and entropy
dependence of isothermal and adiabatic incompressibility of infinite
symmetric nuclear matter with a simple Skyrme interaction.
In Sect.III and IV we will consider the contributions of effective mass
and excluded volume. In Sect.V we consider the incomprssibility of asymmetric 
finite nuclear system which has surface energy, symmetric energy and Coulomb
energy contribution. The conclusion is given in Sect.VI.

\section{General Considerations}

 First, we define a quantity $\kappa$, incompressibility coefficient, as
\begin{eqnarray}
  \kappa = k_F^2 \frac{d^2(E/A)}{d k_F^2} = 9 \rho^2 \frac{d^2(E/A)}{d\rho^2}
         = 9 V^2 \frac{d^2(E/A)}{dV^2} = R^2 \frac{d^2(E/A)}{d R^2}.
                    \label{kappadef}
\end{eqnarray}
This quantity is evaluated at the saturation density $\rho_0$
where $E/A$ has a minimum.
The giant monopole resonance energy is related to $\kappa$
and this relation is $E_0 = \sqrt{\frac{\hbar^2 \kappa A}{m \bra r^2\ket}}$.
If the temperature is kept constant in the above derivatives we have the 
isothermal incompressibility $\kappa$, and if the entropy is held fixed,
the result is the adiabatic incompressibility $\kappa_Q$.
This $\kappa$ is related
to the second order Taylor expansion coefficient of $E/A$ with $\rho$ or
$E/A= a + b (\rho-\rho_0)^2/2$. Thus $\kappa = 9 \rho_0^2 b$.
The quantity $\kappa$
defined above is not the isothermal compressibility defined in thermal 
physics as $K = -(\frac{1}{V}) ({dV}/{dP})_T$ with $T$ held fixed and 
here $P$ is the pressure.
Since $P=-dF/dV|_T$, we
have $K = (\frac{1}{V}) \left(1/\frac{d^2 F}{dV^2}\right)_T$. 
At $T=0$, $F = E-T S = E$, and thus $\kappa = 9/(\rho_0 K)$.
This reciprocal connection between $K$ and $\kappa$ is no longer true 
at finite $T$.  
A reciprocal relation would exist  between $K$ and a $\kappa$ defined by 
an equation similar to Eq.(\ref{kappadef}), but with energy $E$ replaced 
with the free energy $F$.
The $K$ can be connected to the density fluctuations through the equation
 $\langle\rho^2\rangle - \langle\rho\rangle^2 = (\langle\rho\rangle^2/V) T K$
 \cite{huang}.  
Near a first order phase transition and at its critical point, 
the $K$ can become very large since small changes in the pressure can produce
large changes in the volume.  Large density fluctuations at a critical point 
give rise to the phenomena of critical opalescence where the presence of 
droplets of all sizes scatter light.  
In the limit of an ideal gas $K = 1/P$. For a Van der Waals gas, 
whose equation of state is $(P+a(N/V)^2) (V-b N) = N T$, 
the isothermal compressibility $K$ will diverge above and 
near the critical temperature $T_c$ as $1/(T-T_c)^\gamma$,
where the critical exponent $\gamma = 1$ which is the mean field value.
Its reciprocal will go to zero. 
The energy per particle for the Van der Waals gas is $E/N=(3/2) T - a(N/V)$
for $V > bN$, while the entropy per particle
is $S/N=\log[(V-b N)/(N \lambda^3)]+5/2$. The $\lambda$ is the 
De Broglie or quantum  wavelength $\lambda=h/(2\pi m T)^{1/2}$.
The energy per particle monotonically decreases with increasing 
density $\rho = N/V$ and is at it's minimum value at the
smallest density determined by the excluded volume $V = b N$. Besides the 
isothermal compressibility, an adiabatic compressibility can be obtained by
keeping the entropy constant.
The adiabatic compressibility is $K_Q=-(\frac{1}{V}) (dV/dP)_S$
with the derivative at constant entropy.
The reciprocal is related to $\kappa_Q$, the adiabatic incompressibility as
$\kappa_Q=9/(\rho_0 K_Q)$. The energy per particle of a Van der Waals gas
can have a minimum with $V$ at constant entropy.
Since the natural variables for energy are entropy and volume from
$dE=T dS- P dV$ variations of the energy with $V$ at constant $S$ bear
a similar relation to variations of the  Helmholtz free energy
with $V$ at constant $T$ where $dF=-S dT - P dV$.
Thus the adiabatic incompressibility of Eq.(\ref{kappadef}) can go to zero
for a Skyrme interaction as we shall see.
The minimum point (also maximum point)  in the
energy occurs at zero pressure since $dE/dV$ at constant entropy is $-P$.
Therefore $E$ at constant $S$ has the same maximum and minimum points
with variations in $V$ or $R$ or density as $F$ at constant $T$
since both derivatives are $-P$ which is set to 0.

\subsection
{Review of Infinite Matter Incompressibility}

    Our mean field discussion is based on a Skyrme interaction. The Skyrme
interaction shares some common features with the above Van der Waals equation
such as an intermediate density attraction (the $a(N/V)^2$ term above)  and a
higher density repulsion (the excluded volume term in Van der Waals). The
excluded volume leads to a much stronger repulsive term than the Skyrme
repulsive term since it involves an expansion to all orders in the density. 
To keep the discussion simple to begin with, we consider uncharged symmetric
nuclear matter with no surface energy terms.  
The Skyrme interaction energy is then 
\begin{eqnarray}
   U/A= -a_0 \rho + a_\alpha \rho^{1+\alpha}.   \label{meanpot}
\end{eqnarray}
The $a_0$ term gives a medium range attraction while the $a_\alpha$ term is 
a short range repulsion.

At $T=0$, the kinetic energy  $E_K/A = (\frac{3}{5}) E_F(\rho)$. 
The coefficients $a_0$ and $a_\alpha$ are fixed to give a binding energy 
per particle $E_B/A=16$MeV at saturation density $\rho = \rho_0 = 0.15$/fm$^3$.
At this density $E_F = \frac{\hbar^2}{2m} (\frac{6\pi^2}{4}\rho)^{2/3} = 35$MeV
and $E_K/A=21$MeV. 
From $E = E_K + U$ and $d(E/A)/d\rho = 0$ at $\rho_0$ the coefficients are
\begin{eqnarray}
 a_0 \rho_0 = b_0 &=& (E_K+E_B)/A + b_\alpha
     = (E_K+E_B)/A+(1/\alpha) (E_K/3+E_B)/A = 37 + 23/\alpha  \nonumber \\
 \alpha a_\alpha \rho_0^{1+\alpha} &=& \alpha b_\alpha
     = (E_K/3+E_B)/A = 23.   \label{coeff}       \\
 (1+\alpha) b_\alpha &=& b_0 - \frac{2}{3} E_K/A   \nonumber
\end{eqnarray}
The incompressibility coefficient $\kappa$ at $T=0$ is then 
\begin{eqnarray}
  \kappa = - 2 E_K/A + 9 (1+\alpha) \alpha a_\alpha \rho_0^{1+\alpha}
         = E_K/A + 9 E_B/A + \alpha (3 E_K/A + 9 E_B/A) = 165 + 207 \alpha .
              \label{kappat0}
\end{eqnarray}
For $\alpha=1/3$, $\kappa=234$MeV at $\rho_0$.
Smaller values of $\alpha$ lead to softer equations of state and lower $\kappa$.
In the limit $\alpha \to 0$, logarithmic terms appear in Eq.(\ref{meanpot})
coming from the presence of a factor
$(x/\alpha) (1-x^\alpha) = (x/\alpha) (1-e^{\alpha \log x}) \to -x\log(x)$. 
The $x=\rho/\rho_0=(R_0/R)^3$ with $R_0^3=A/(\rho_0 4\pi/3)$.
The $\alpha \to 0$ limit is the softest EOS allowed by Eq.(\ref{meanpot}),
and this limit gives from Eq.(\ref{kappat0}) a value of $\kappa = 165$ MeV.
A stiff EOS has $\alpha=1$ and $\kappa=372$ MeV. 
Recent calculations \cite{piekar,soubb,vrete}
have a value of $\kappa = 210 \sim 270$ MeV
and suggest a value of $\alpha=1/3$ in a Skyrme type approach.
A larger range of values of $\kappa$, from $211 \sim 350$ MeV,
were reported in Ref.\cite{lalazi}.
Because of these uncertainties in $\kappa$ from more realistic forces,
we will present results for various values of $\alpha$,
from $\alpha\approx 0$ to $\alpha=1$.

\subsection
{Corrections to the Infinite Matter Result}
\label{sectsk}

 If the corrections to the infinite nuclear matter incompressibility from
finite size terms or non-zero temperature or entropy terms are small,
then these corrections can be obtained by using the following method.
Let $E_0(R)$ be the nuclear matter energy per particle EOS 
which has a minimum at $R_0$ and an incompressibility $\kappa_0$.
If we add to this a term $E_x(R)$,
so that $E(R)=E_0(R)+E_x(R)$ then the minimum shifts to a new point
$R_m = R_0 + \Delta R_x$.
The new minimum and $\kappa$ can be found by making a Taylor
expansions around of $R_0$. The shift is given by
\begin{eqnarray}
  \Delta R_x/R_0 = - R_0 (d E_x(R)/dR) /\kappa_0   
                   \label{delrx}
\end{eqnarray}
with the derivative evaluated at $R_0$
since $R^2 \frac{d^2 E/A}{dR^2}|_{R_m} \approx R^2 \frac{d^2 E/A}{dR^2}|_{R_0}$
for small $\Delta R_x$.
The new $\kappa$ is
\begin{eqnarray} 
  \kappa &=& R^2 \frac{d^2 \left({E(R)}/{A}\right)}{dR^2} 
          = \kappa_0 + R^2\left(\frac{E_x(R)}{A}\right)''
             - 2 R \left(\frac{E_x(R)}{A}\right)' 
             - R\left(\frac{E_x(R)}{A}\right)' \frac{Sk}{\kappa_0}
                   \label{kappaex}      \\
  Sk &=& R^3 \left.\frac{d^3 \left({E_0(R)}/{A}\right)}{dR^3}\right|_{R_0}
        = - k_F^3 \frac{d^3 \left({E_0(R)}/{A}\right)}{d k_F^3}
                   \label{skewness}
\end{eqnarray} 
The various quantities are evaluated at $R_0$ and 
each $'$ represents one derivative wrt $R$.
Corrections to $\kappa$ involving skewness $Sk$ or third derivative of 
the energy were pointed out in Ref.\cite{blaizot,pears}.
However note here that the skweness $Sk$ is defined with opposite sign 
from others.
Ellis et al \cite{ellis} used the correlation between 
compression modulus and skewness coefficient to examin
the implications in a relativistic Hartree-Fock approximation
where the $E_x$ is the Coulomb interaction.
The above expression is a modified  version of their result.
When two small correction terms are present then the above result is
easily generalized and each adds its own separate contribution.
If the corrections are large then the above approximate expression 
will not be very accurate and it is best to calculate the shift
and $\kappa$ exactly.
We will use these results for the shift and $\kappa$ to find simple
expressions for $\kappa$ for the general case of arbitrary $\alpha$.
Eq.(\ref{kappat0}) gives an expression for the incompressibility
at $T=0$ in terms of $E_K$ and $E_B$. 
This will be $\kappa_0 = \kappa_0(\alpha)$ in Eq.(\ref{delrx})
and Eq.(\ref{kappaex}).
The skewness $Sk=R^3 (E_0(R)/A)'''$,
in terms of $E_K$, $E_B$ and $\alpha$, is
\begin{eqnarray}
  Sk(\alpha) = - 27 (E_B/A) (3 + 4\alpha + \alpha^2)
                 - (E_K/A) (11+36\alpha+9\alpha^2)
             =-3 (509 + 828 \alpha + 207 \alpha^2).
               \label{skew}
\end{eqnarray}
The last equality in Eq.(\ref{skew}) follows when $E_K=21$MeV and $E_B=16$MeV.
Comparing this last equation with 
the $\kappa$ of Eq.(\ref{kappat0}) we see that the ratio of $Sk/\kappa_0$
is of the order of 10 and somewhat insensitive to $\alpha$.
In particular for $\alpha=0$, 1/3 and 1 this ratio
is $-9.25$, $-10.36$ and $-12.45$.
We will apply this procedure or exact solutions to evaluate the
contribution of surface, Coulomb, symmetry energy,
temperature and entropy  to the nuclear incompressibility.
For the particular case $\alpha = 1/3$, exact solutions are also easy
to obtain in several situations which lead to a quadratic equation for 
the equilibrium point $R$ when $E$ is written in terms of $R$.

\subsection
{Non-zero Temperature and Entropy Considerations}

 At non-zero $T$ and with $T \ll E_F$, the kinetic energy is
 $E_K/A = (\frac{3}{5}) E_F + (\frac{\pi^2}{4}) T^2/E_F$ \cite{huang,sjllg}. 
The energy per particle is
\begin{eqnarray}
   E/A = 21 x^{2/3} + \left(\frac{\pi^2}{140}\right) \frac{T^2}{x^{2/3}}
               - b_0 x + b_\alpha x^{1+\alpha}.  \label{epa}
\end{eqnarray}
This equation can not be used for very small $x$ since the condition
 $T \ll E_F$ will fail.
We can use the procedure of Sect.\ref{sectsk} \cite{prlsub}. 
 Taking $E_x(R,T)= 0.0517 T^2 R^2/A^{2/3}$ in Eq.(\ref{kappaex}) and
(\ref{skewness}), 
we obtain the following expression for $\kappa$ and $x_m$:
\begin{eqnarray}
  \kappa(\alpha) &=& \left. R^2 \frac{d^2(E(R)/A)}{dR^2} \right|_T
       = \kappa_0(\alpha)
            - 0.1034 T^2 R_0^2/A^{2/3} \{Sk(\alpha)/\kappa_0(\alpha)+1\}
           \label{kappalt}  \\
  x_m(\alpha) &=& \rho_m/\rho_0 = (R_0/R_m)^3
       = (1 + \Delta R_x/R_0)^{-3} = (1 - 0.1034 T^2/A^{2/3})^{-3}
\end{eqnarray}
Note here that $x_m > 1$, i.e., the saturation density is larger than $\rho_0$
in an isothermal process.
Since $A/(4\pi R_0^3/3)= \rho_0=0.15/$fm$^3$ and 
thus $R_0^2/A^{2/3}=0.734$fm$^2$
we have the following final simple results:
\begin{eqnarray}
   \kappa(\alpha=0) &=& 165 + 1.16 T^2       \nonumber \\
   \kappa(\alpha=1/3) &=& 234 + 1.32 T^2     \nonumber \\
   \kappa(\alpha=1) &=& 373 + 1.61 T^2      \label{kappalt1}
\end{eqnarray}
From Eq.(\ref{kappalt1}) 
we see that the first term is very sensitive to $\alpha$ but the finite
temperature correction is somewhat insensitive to $\alpha$ \cite{prlsub}.
Since $Sk/\kappa_0$ is insensitive to $\alpha$ with a value of about $-10$,
the second term does not scale with the first term in Eq.(\ref{kappalt1})
as $\alpha$ changes.
At $T=7.5$MeV the $T$ dependent corrections in Eq.(\ref{kappalt1})
are 65, 74 and 91MeV 
while the first terms are 165, 234 and 373 MeV
for $\alpha=0$, $\alpha=1/3$ and $\alpha=1$.    
These results are very close to the exact solution result 
of Ref.\cite{prlsub}.  
The exact result is given by
  \begin{eqnarray}
    \kappa(T) = -42 x_m^{2/3} + 0.705 T^2/x_m^{2/3} + 9 \alpha (1+\alpha) b_\alpha x_m^{1+\alpha}  
  \end{eqnarray}
 At $T=2.5$, 5,  and 7.5MeV,  and for $\alpha=1/3$, the values of $\kappa$ are 242, 265 and 
 302MeV, respectively.  
 The value of $\alpha=1/3$ and $T=0$ is
 also useful. It leads to an energy whose minimum can be obtained from a quadratic
 equation in the radius using $\rho=A/(4\pi R^3/3$). Namely $E(R)/A$ is
  \begin{eqnarray}
   E(R)/A =  21\left(\frac{R_0}{R}\right)^2 - 106\left(\frac{R_0}{R}\right)^3
                   + 69\left(\frac{R_0}{R}\right)^4
          = 28.626 \frac{A^{2/3}}{R^2} - 168.7 \frac{A}{R^3}
                   + 128.2 \frac{A^{4/3}}{R^4}
    \label{era}
 \end{eqnarray}
 At finite $T$ a term $(\pi^2/140) T^2 (R/R0)^2=0.0517 T^2 R^2/A^{2/3}$
 is added to the right side of Eq.(\ref{era}). 
 When $T$ is replaced with entropy per particle $S/A$ then this $T$ dependent
 term becomes
 $4.836 (S/A)^2 A^{2/3}/R^2$ since $S=(\pi^2/2) T A/E_F$ at low $T$.
 This $S/A$ term can simply be added to the 
 first term on the right side of Eq.(\ref{era}) since both have the
 same $A^{2/3}/R^2$ dependence.
 In the limit $\alpha \to 0$,
  $x_m$ satisfies the equation $14 (x^{2/3}-x) + 23x\log x = (\pi^2/210) T^2/x^{2/3}$.
 At $T=0$, 2.5, 5, 7.5, the $x_m=1$, 1.0157, 1.0598 and  1.1251,
 while $\kappa = -42x^{2/3} + 0.705 T^2/x^{2/3} + 207x$ 
 has values 165, 172, 193, 224 MeV, respectively.

 At fixed entropy, the second derivative of $E(R)/A$ has a very different
behavior than at fixed $T$. Namely, it decreases with $R$ for a given $S/A$.
This can easily be seen by noting that $E_x(R,S)=4.836(S/A)^2 A^{2/3}/R^2$
compared to $E_x(R,T) =0.0517T^2 R^2/A^{2/3}$.
Here we used $S = \frac{\pi^2}{2} T A /E_F$ at low $T$.
Because $R$ now is in the denominator, the first derivative is negative
and $\kappa_Q$ will decrease
\begin{eqnarray}
  \kappa_Q &=& \left. R^2 \frac{d^2(E(R)/A)}{dR^2} \right|_S
       = \kappa_0(\alpha)
        + 9.672(S/A)^2 A^{2/3}/R_0^2\{Sk(\alpha)/\kappa_0(\alpha)+5\} 
              \label{kappals}  \\
  x_m &=& (1 + 9.672(S/A)^2 A^{2/3}/R_0^2)^{-3}
\end{eqnarray}
Note here that $x_m < 1$, i.e., the saturation density is lower than $\rho_0$
in an adiabatic process in contrast to an isothermal process.
We have the following final results:
\begin{eqnarray}
   \kappa_Q(\alpha=0) &=& 165 - 30 (S/A)^2       \nonumber \\
   \kappa_Q(\alpha=1/3) &=& 234 - 38 (S/A)^2     \nonumber \\
   \kappa_Q(\alpha=1) &=& 373 - 53 (S/A)^2       \label{kappals1}
\end{eqnarray}
in MeV.

 At higher $T$, the nearly degenerate Fermi gas kinetic energy term is 
replaced by a virial expansion in $\rho \lambda^3$, where $\lambda$ is 
the quantum wavelength.  
Namely,  $E_K/A= (3/2) T (1+\sum{c_n (\rho \lambda^3/4)^n})$
where the sum over $n=1$, 2, 3, $\cdots$ has coefficients:
 $c_1=1/2^{5/2}=0.177$, $c_2=(1/8-2/3^{5/2})=-3.3 \times 10^{-3}$,
 $c_3= 1.11 \times 10^{-4}$, $\cdots$ \cite{huang,sjllg}.
These coefficients come from antisymmetrization effects only.
The role of clusters in the virial expansion is discussed in Ref.\cite{prlsub}.
Since the $c_n$'s become small rapidly, we will keep terms up to $c_2$.
Then $\kappa$ is given by:
\begin{eqnarray}
  \kappa(T) = - (2196/T^2) x_m^2
       + 9 \alpha (1+\alpha) b_\alpha  x_m ^{1+\alpha}  \label{kappat} 
\end{eqnarray}
The $x_m$ is again the minimum of  $E/A$, but now evaluated with the new
kinetic energy. The $x_m$ is affected by both the $c_1$ and $c_2$ terms.
A limiting value of $\kappa$ can be obtained by taking $T$ very large where
an ideal gas $E_K/A=(3/2)T$ leads to very simple results \cite{prlsub}. 

  We also
investigate the case of constant entropy in the ideal gas limit
using the Sackur-Tetrode law \cite{sakur}: $S/A=5/2-\ln(\lambda^3 \rho/4)$.
This law connects $T$ to $\rho$ or $V$ as $T=C_S \rho^{2/3}$.
Here $C_S = [2\pi(\hbar c)^2/(mc^2)] \exp[(2/3) (S/A)-5/3]$.
Then $E_K/A= (3/2) C_S \rho^{2/3}
         (1 + \rho^{2/3} \sum{c_n ([2\pi\hbar^2/m C_S]^{3/2})^n})$.
The resulting $E(R)/A$ has a structure similar to the result for a degenerate 
Fermi gas since both have a $\rho^{2/3}$ dependence for the kinetic energy
term but with different coefficients.
This feature and a similar result at lower $T$ suggests a duality 
in the energy per particle EOS at constant entropy and its associated 
$\kappa_Q$ and the $T=0$ EOS and its associated constant $T$ $\kappa$ 
\cite{prlsub}.

 It should be noted that the $x_m$'s for the constant $T$ case are 
 usually close to 1 except at high $T$, and greater then 1, so that the
 minimum of the energy occurs at a density that is greater than $\rho_0$.
 This feature is not the case for the
 Helmholtz free energy $F = E-T S$.  Since $P=-dF/dV$,  $F$ is minimum when $P=0$.
 At low $T$,  the entropy $S=(\pi^2/2) T A/E_F$, and thus $F_K=E_K-T S$, the part of the
 Helmholtz free energy coming from the kinetic energy part of $E$,
 is $F_K/A=(3E_F/5) (1- (5\pi^2/12) (T/E_F)^2)$. 
 The complete $F/A=F_K/A+U/A$, where $U/A$ is given by Eq.(\ref{meanpot}).  
 The  free energy per particle at low $T$ can be obtained from Eq.(\ref{epa})
 by just changing the sign of the $T^2$ term:
  $F/A=21 x^{2/3} - (\pi^2/140) T^2/x^{2/3} -b_0 x + b_\alpha x^{1+\alpha}$.
 As with the $E/A$ equation, this $F/A$ equation is only valid when $T \ll E_f$.
 For higher $T$, we use the virial expansion for $E_K$ involving the coefficients $c_n$ mentioned
 above. Similarly,
 the $T S/A = (5/2) T - T \ln(\lambda^3 \rho_0 x/4)
        + c_1 (\lambda^3 \rho_0 x/4)/2+ \cdots$ \cite{huang,sjllg}.
 We note that the minimum of the free energy $F(T,V)$
 and energy $E(T,V)$ at constant $T$ don't coincide except at $T=0$. 
 However, the minimum of free energy $F(T,V)$ at constant $T$ and
 the minimum of energy $E(S,V)$ at constant $S$  are at $P=0$.
 The minimum of $F$ is at 
 subnuclear density, or $x_m<1$, at non zero $T$. 
 For $\alpha=1/3$ and at $T=0$, 2.5, 5, 7.5 MeV
 the $x_m=1$, 0.988, 0.951, 0.878 and the curvature function
 $\kappa = 9 x_m^2 (d^2(F/A)/dx_m^2) = 234$, 225, 199, 150 MeV, respectively.
 The $F$ also has a maximum at even smaller $x$
 than its minimum point. The pressure $P = \rho x d(F/A)/dx = \rho_0 x^2 d(F/A)/dx$ and
 is $P = 2.1x^{5/3}+0.00705T^2 x^{1/3}-0.15 b_0 x^2+0.15 (1+\alpha) b_\alpha x^{2+\alpha}$
 for $T \ll E_F$. 
 At high $T$ and/or small $x$,
 the $P=\rho_0 x T \{1+\sum c_n (\lambda^3 \rho_0/4)^n x^n\}
   - 0.15 b_0 x^2+0.15 (1+\alpha) b_\alpha x^{2+\alpha}$.
 The spinodal line is obtained by setting $dP/dV=0$
 or $d^2(F/A)/dV^2 = 0$. The critical point occurs at the top of the spinodal
 line where $d^2 P/dV^2=0$ also.
 Similar remarks also apply to $E(S,V)$ with $S$ held
 constant and for different values of $S$.
 The specific heat at constant volume is also easily obtained from the above results.
 At low $T$, $C_V$ is linear in $T$ and is $(\pi^2/2) T/E_F$.
 The behavior of $C_V$ at higher $T$
 can be obtained from the virial expansion of $E_K/A$ given above. 
 At very high $T$ the $C_V \to 3/2$.
 The $T=0$ limit where $C_V=0$ is connected to this very high $T$ limit of
 3/2 in a monotonic way with increasing $T$ with no peak or cusp in $C_V$.
 This is to be compared with the strong peak found in $C_V$
 when inhomogeneities are included \cite{dasmek}.
 In this latter case, cost functions arising from the surface energy associated with
 the formation of clusters are responsible for the very large peak as mentioned.
 These calculation \cite{dasmek} were done at a freeze out density of about 1/3
 nuclear matter density.
 At high density, i.e., above nuclear matter, heated matter may first form
 bubbles in a uniform background of liquid. The formation of a bubble has
 associated with it a surface energy. 
This aspect will be considered in a future paper.

\section{
 Effective Mass Effects}

  An effective mass $m^*$ changes the kinetic energy term.
If we parameterize it as $m^*/m=1/(1+r(\rho/\rho_0)^\gamma)$,
then at zero temperature the $E/A$ will simply read
  $E/A = (E_K/A) x^{2/3} + r (E_K/A) x^{2/3+\gamma}
      - b_0 x + b_\alpha x^{1+\alpha}$.
The $E_K/A=(3/5) E_F = 21$MeV, the value with the unrenormalized mass, 
and the effective mass correction appears in
the additional second term with a $\gamma+2/3$ power dependence.
The $r$ is a constant and has a value of 1/2 when $m^*/m=2/3$ at $\rho_0$.
If $\alpha=\gamma-1/3$, then this additional effective mass term
could be simply added to the $b_\alpha$ term.
Since the coefficients are determined to give the correct binding at 
$\rho=\rho_0$ or $x=1$ where $E/A$ has a minimum, the $b_\alpha$
coefficient must be adjusted also.
The final incompressibility is still given by Eq.(\ref{kappat0})
with $\alpha=\gamma-1/3$. 
If we parametrize $m^*/m$ with $\gamma = 1$
then for $\alpha=2/3$ the effective mass term can be added to 
the $b_\alpha$ term and the final incompressibility is now 
given by Eq.(\ref{kappat0}) with $\alpha=2/3$.
Similarly if we parametrize $m^*/m$ with $\gamma = 2/3$
then for $\alpha=1/3$ the effective mass term can be added to 
the $b_\alpha$ term and the final incompressibility is now 
given by Eq.(\ref{kappat0}) with $\alpha=1/3$.
If $\gamma = 1/3$, the $r$ dependent effective mass term can be added to 
$b_0$ term  and there is no effect on $\kappa$ and $Sk$.
For general $\alpha$ and $m^*/m$: 
 $\alpha b_\alpha = (E_B/A) + (1-[3\gamma-1]r) (E_K/A)/3$
and $b_0 = (1+r) (E_K/A) + (E_B/A) + b_\alpha$ and
\begin{eqnarray}
 \kappa_0(\alpha,r) &=& 9 E_B/A + (1+[3\gamma-1]^2 r) E_K/A
             + \alpha (9 E_B/A + 3(1-[3\gamma-1]r)E_K/A)   \nonumber \\
     &\to& 3 (55 + 69 \alpha + 7 r [3\gamma-1] ([3\gamma-1]-3\alpha)) 
\end{eqnarray}
For $\gamma=1/3$ or for $\alpha=\gamma-1/3$ this $\kappa$ is $r$ independent 
for reasons already given.
When we parametrize $m^*/m$ with $\gamma=1$,
for $\alpha=1/3$, $\kappa = 234+42r$ and for $r=1/2$, $\kappa = 255$MeV.
For $\alpha=1$, $\kappa = 372 - 42r \to 351$MeV at $r=1/2$.
The skewness is also affected by the effective mass:
\begin{eqnarray}
    Sk &=& -27 (E_B\!/\!A) (3 + 4\alpha + \alpha^2)
          - (E_K\!/\!A)(11 + 36\alpha + 9\alpha^2
          + r [3\gamma-1] ([3\gamma-1] - 3\alpha)([3\gamma+11] + 3\alpha))
             \nonumber \\
       &\to&  -3(509+828\alpha+207\alpha^2)
              - 21 r [3\gamma-1] ([3\gamma-1]-3\alpha)([3\gamma+11]+3\alpha)
\end{eqnarray}
For $\gamma=1/3$ or for $\alpha=\gamma-1/3$ this is independent of $r$ and 
for $\alpha<\gamma-1/3$, $Sk/\kappa_0$ decreases with $r$;
for $\alpha > \gamma-1/3$, $Sk/\kappa_0$ increases with $r$.

  The effective mass also appears in the temperature dependent part of the 
energy per particle, modifying the additional term $E_x(R,T)$ to now
be given by $E_x(R,T)=0.0517(T^2 R^2/A^{2/3}) (1/(1+r (R_0/R)^{3\gamma}))$.
The last factor in this
expression is the change the effective mass produces.
The result of Eq.(\ref{kappaex}) can be used
to calculate the new $\kappa$. This leads to the following equation:
\begin{eqnarray}
 \kappa &=& \kappa_0(\alpha,r)    \nonumber  \\  & &
   - 0.0705 T^2 \left[
     \frac{(2 + ([3\gamma-3][3\gamma+2]+10)r - [3\gamma-1][3\gamma+2]r^2)}
                   {(1+r)^3}
       + \frac{(2+[3\gamma+2]r)}{(1+r)^2}
             \left(\frac{Sk(\alpha,r)}{\kappa_0(\alpha,r)}\right) \right]  
\end{eqnarray}
For the case of $\gamma=1$,
at $T=7.5$MeV and $r=1/2$: $\kappa=283$MeV for $\alpha=0$,
335MeV for $\alpha=1/3$,
388MeV for $\alpha=2/3$ and 442MeV for $\alpha=1$.
The $T=0$, $r=1/2$ values of $\kappa$ are 207, 255, 303, and 351 MeV 
and values of $Sk$ are 2115, 2739, 3459, and 4275 MeV for each of these 
respective $\alpha$ values.    
For a fixed entropy,
 $E_x(R,T)=4.836(S/A)^2 (A^{2/3}/R^2) (1/(1+r (R_0/R)^{3\gamma}))$ 
and thus
\begin{eqnarray}
 \kappa_Q &=& \kappa_0(\alpha,r)          \nonumber  \\  & &
    + 3.548 (S/A)^2 \left[
     \frac{(10 - ([3\gamma-2][3\gamma+9]-2)r + [3\gamma-2][3\gamma-5]r^2)}
                 {(1+r)^3}
       + \frac{(2-[3\gamma-2]r)}{(1+r)^2}
             \left(\frac{Sk(\alpha,r)}{\kappa_0(\alpha,r)}\right) \right]  
\end{eqnarray}

At high $T$ with effective mass,
 $E_K/A = (3/2)T(1 + \sum_n c_n (\rho(1 + r(\rho/\rho_0)^\gamma)^{3/2}
      \lambda^3/4)^n)$
due to the mass dependence in the De Broglie wavelength $\lambda$.
Due to the density dependence of the effective mass the $c_1$ term
contributes directry to incompressibility now.
Then $\kappa$ is, up to $c_1$ term, given by:
\begin{eqnarray}
 \kappa(T) = \frac{1505}{T^{1/2}} \frac{x_m}{(1+r x_m^\gamma)^{1/2}} 
         r x_m^\gamma \frac{3}{2} \gamma 
        \left(1 + \gamma + (1 + \frac{3}{2} \gamma) r x_m^{\gamma}\right)
       + 9 \alpha (1+\alpha) b_\alpha  x_m ^{1+\alpha}  
\end{eqnarray}
The $x_m$ is again the minimum of $E/A$, but now evaluated with the new
kinetic energy. 
At fixed entropy 
 $E_K/A= (3/2) C_S \rho^{2/3} (1 + \sum{c_n ([2\pi\hbar^2/m C_S]^{3/2} 
    (1 + r(\rho/\rho_0)^\gamma)^{3/2})^n})$.
The lowest order term having effective mass dependence is the $c_1$ term.
Up to this term,
\begin{eqnarray}
 \kappa_Q(S) &=& -3 C_S \rho_0^{2/3} x_m^{2/3} 
    + \frac{3}{2} C_S \rho_0^{2/3} \left(\frac{2\pi\hbar^2}{m C_S}\right)^{3/2}
          \frac{c_1 x_m^{2/3}}{(1+r x_m^\gamma)^{1/2}}
                    \nonumber  \\  & &   \hspace{4em}  \times
        \left(-2 + \frac{(27\gamma^2+9\gamma-8)}{2} r x_m^\gamma
          + \frac{(9\gamma+4)(9\gamma-2)}{4} r^2 x_m^{2\gamma}\right)
       + 9 \alpha (1+\alpha) b_\alpha  x_m ^{1+\alpha}  
\end{eqnarray}

\section{
 Excluded Volume Effects}  

  Excluded volume effects appear in the Van der Waals approach and modify the
 adiabatic incompressibility, leading to an equation of state of the form,
for a given $S$,
\begin{eqnarray}
    E/A = (3/2) C_S [\rho/(1-b \rho)]^{2/3} - a_0 \rho    \label{ecs}
\end{eqnarray}
where $C_S = [2\pi(\hbar c)^2/(mc^2)] \exp[(2/3) (S/A)-5/3]$
which connects $T$ to $\rho$ or $V$ as $T=C_S \rho^{2/3}$ for ideal gas.
An expansion of
 $1/(1-b\rho)^{2/3} = 1 + (2/3)(b\rho) + (5/9)(b\rho)^2+ \cdots$ leads to 
an infinite series of repulsive terms with various powers of the density.
The numerator has $\rho^{2/3}$ so that additional terms such as
$\rho^{5/3}$, $\rho^{8/3}$, $\rho^{11/3}$, ... 
are generated in this expansion.
These terms collectively give rise to a much high incompressibility.
If we had truncated the series at $\rho^{5/3}$, then the
resulting EOS would be similar to a Skyrme with $\alpha=2/3$.
A Skyrme with $\alpha=2/3$ has  an incompressibility of  303MeV.

To explore higher density repulsive terms in a $T=0$ Skyrme parametrization,
we can use the duality bewteen the EOS at constant entropy and the EOS 
at zero $T$ \cite{prlsub}.
Thus we take the specific form of Eq.(\ref{ecs}) with $(3/2) C_S$ replaced 
by a constant parameter $a_K$ at zero $T$ as a very simple model
for studying excluded volume like effects in our $E/A$.
We can fix the coefficient $a_0$ and $a_K$
to give a binding energy per particle  of $16$MeV at $\rho=0.15$/fm$^3$:
$b_K = a_K \left(\frac{\rho}{1-b\rho}\right)^{2/3} 
       = 48 \left(\frac{1-b\rho}{3b\rho-1}\right)$
and $b_0 = a_0\rho = \frac{32}{(3b\rho-1)}$.
The resulting $\kappa$ is
\begin{eqnarray}
  \kappa &=& 96 (6b \rho-1)/((3b \rho-1) (1-b \rho))
\end{eqnarray}
This $\kappa$ has a minimum value of 754MeV at $b = 3.60$.
Alternatively if we fix $a_K$ to be the $T=0$ Fermi kinetic
energy coefficient, $a_K \rho_0^{2/3} = 21$MeV, and determine $a_0$
and $b$ by having correct binding energy at saturation 
we get $b = 3.61$ and an incompressibility of 754MeV.
The two methods result in very similar values for $b$ and $\kappa$.
A high incompressibility of 754MeV is equivalent to having a
single Skyrme repulsive term with $\alpha = 2.85$ from Eq.(\ref{kappat0}).
This high incompressibility is unreasonable but it
can be moderated by having higher order attractive terms in the density which,
for example, can come from higher order attractive correlations of
nucleons \cite{prlsub}.
The skewness is given by
\begin{eqnarray}
 Sk = 192 \left(\frac{b\rho-1}{3b\rho-1}\right)
       \left[6 + 31 \left(\frac{b\rho}{1-b\rho}\right)
               + 45 \left(\frac{b\rho}{1-b\rho}\right)^2
               + 20 \left(\frac{b\rho}{1-b\rho}\right)^3\right]
     + \frac{1920}{3b\rho-1}
\end{eqnarray}
The effects of temperature and entropy dependences coming from a kinetic 
energy can be obtained from Eqs.(\ref{kappalt}) and (\ref{kappals}) 
with the $\kappa_0$ and $Sk$ given above.

\section{
Asymmetric and Finite Systems} 

   For $N$, $Z$ asymmetric systems with Coulomb and finite size effects 
included, the $T=0$ $\kappa$ is reduced \cite{zamick,blaizot}. 
This was qualitatively explained in Ref.\cite{zamick} as a 
reduction in the binding energy term in Eq.(\ref{epa}). 
Detailed calculation done in Ref.\cite{blaizot} confirmed this feature.
In this part of the paper we will present some very simple 
equations for the contribution of surface, symmetry and surface symmetry,
and Coulomb and surface Coulomb.  
The surface, symmertry energy and Coulomb terms each reduce the binding 
energy per particle from 16MeV to 8MeV. 
For light nuclei around $A=27$, the surface energy/$A \sim 6$MeV, 
the symmetry energy$/A$ $\sim$ 0MeV and the Coulomb energy/$A$ $\sim$ 1.5MeV.
For heavy nuclei around $A=216$, the above numbers are changed to $\sim$3MeV,
$\sim$1MeV and $\sim$4MeV for the surface, symmetry and Coulomb terms, 
respectively.
We will find a somewhat similar pattern of relative importance in $\kappa$,
but the surface term will be dominant even for large nuclei. The Coulomb is 
of next importance followed by the symmetry term for nuclei near the valley 
of beta stability. 
Surface symmetry and surface Coulomb terms will also be included.

\subsection{
Surface Contribution}

 First, let us investigate the role of finite size or surface
correction, before considering the full complexity of the problem with
Coulomb and symmetry energy terms included in the energy.  
The surface energy can be studied by using 
 $\rho(r) = \rho_c/(\exp[(r-R)/a]+1)$ 
as a density profile.
Here $\rho_c$ is the central density and $a$ is a parameter 
that determines the surface thickness. 
This density can then be used in a Skyrme energy density functional:
  $E(\rho) = a_K \rho^{5/3} - a_0 \rho^2 + a_\alpha \rho^{2+\alpha}$.
The $a_K$ is the kinetic energy constant 
and is taken as $a_K \rho_0^{2/3}=(3/5) E_F=21$MeV. 
The $a_0$ and $a_\alpha$ are given above. 
The energy per particle is the integral of $E(\rho$) over all space 
divided by $A$. 
The resulting integrals can be simplified by an integration by parts and
very accurately approximated by noting that the derivative of the Wood-Saxon
profile gives a narrow peak of width $a$ in the vicinity of $R$.
A substitution $y=(r-R)/a$ and a change in the lower limit of integration 
from $-R/a$ to $- \infty$ gives a very accurate approximation to 
these integrals when $R \gg a$, as is usually the case.
As an example, the energy per particle from the $a_0$ attractive interaction 
term is $(a_0/A) (4\pi R^3/3) \rho_c^2[1-3(a/R)+\pi^2(a/R)^2-\pi^2(a/R)^3]$. 
The central density $\rho_c$, $R$ and $a$ are also connected since the 
integral of $\rho$ over all space must give $A$.
Using the integration procedure just mentioned, the condition
 $(4\pi R^3/3) \rho_c(1+\pi^2(a/R)^2)=A$ is obtained. 
Note that in this last result the linear and cubic terms in $a/R$ are 
absent in the constraint equation. 
This is not the case for the integrals involving the kinetic energy density, 
the $a_0$ and the $a_\alpha$ terms. 
Each of these has a form involving surface thickness corrections that
is $(1-c_1(a/R)+c_2(a/R)^2-c_3(a/R)^3)$. 
For the kinetic energy term $c_1=2.2769$, $c_2=9.1046$, $c_3=7.805$. 
For the $a_0$ term $c_1=3$, $c_2=c_3=\pi^2$, and for 
the $a_\alpha$ term, for $\alpha=1/3$, $c_1=3.585$, $c_2=10.819$ 
and $c_3=11.642$,
for $\alpha=1$, $c_1=9/2$, $c_2=\pi^2+3$ and $c_3=3\pi^2/2$.
The energy per particle can be rewritten as:
\begin{eqnarray}
   E(R)/A = 28.626 C_K (A^{2/3}/R^2) - (58.89+36.61/\alpha) C_0 (A/R^3)   
             + (23/\alpha) C_\alpha (A/(0.8565R)^3)^{\alpha+1}     \label{eras}
\end{eqnarray}
The $C_K$, $C_0$ and $C_\alpha$ are the surface correction factors,
each involving their corresponding $c_1$, $c_2$, $c_3$ terms as
 $[1-c_1(a/R)+c_2(a/R)^2- c_3(a/R)^3]/[1+\pi^2 (a/R)^2]^j$,
where the denominator arises from the 
constraint condition between $R$, $\rho_c$ and $A$.
The $j=5/3$ for the kinetic term, $j=2$ for the $a_0$ term 
and $j=\alpha+2$ for the $a_\alpha$ term.
For $a/R \ll 1$, the surface energy can be approximated by keeping linear 
terms in $a/R$ to give
\begin{eqnarray}
   E_{diff}(R)/A &=& -(a/R) (21\times 2.2769 (R_0/R)^2
                   - (37+23/\alpha) 3 (R_0/R)^3
                   + (23/\alpha) (3.585) (R_0/R)^{3\alpha+3})  
                 \nonumber \\
 &=& - (a/R) (65.18A^{2/3}/R^2-506.11A/R^3+459.65A^{4/3}/R^4)  
             \label{esuf}
\end{eqnarray}
for $\alpha=1/3$.
This $E_{diff} \to 0$ as $a \to 0$ and $E_{diff} = 10.36 A^{2/3}$ MeV 
at $a=0.53$fm and $\rho_0 = 0.15$fm$^{-3}$.
To have correct surface energy of $\sim 20 A^{2/3}$ MeV,
we also consider a gradient dependent energy density of
 $B (\nabla\rho)^2 = -B \rho\nabla^2\rho$.
With the Wood-Saxon density profile this becomes
\begin{eqnarray}
  E_{grad}/A &=& B \frac{\rho_c}{2 Ra}
        \frac{[1 + (a/R)^2(\pi^2/3 - 2)]}{[1 + \pi^2 (a/R)^2]^2}
     \approx B \frac{\rho}{2 R a} = B \frac{3}{8\pi a} \frac{A}{R^4}
      = \frac{3 B}{8\pi a^2} \frac{a}{R} \frac{A}{R^3}
          \nonumber \\
     &=& \frac{3 B}{8\pi a^2} \frac{a}{A^{1/3}}
        \left(\frac{4\pi\rho_0}{3}\right)^{4/3} \left(\frac{R_0}{R}\right)^4
\end{eqnarray}
The approximate form of this term is for $a/R \ll 1$ and is the same 
form as the $a_0$ term in $E_{diff}$.
The total surface energy then becomes
\begin{eqnarray}
 E_{surf} = 4\pi R^2 \sigma(\rho) = 4\pi r_0^2 A^{2/3} \sigma(\rho)
       = E_{diff} + E_{grad}
\end{eqnarray}
The surface energy has a contribution from the gradient term that
varies with $a$ as $1/a$ while the diffuseness term varies directly
with $a$.
Note that as the surface gets sharper or $a \to 0$, the gradient
term becomes very large and the decrease in binding from the
surface diffuseness becomes small.
Alternatively, as the surface thickness $a$ becomes large,
the gradient term is reduced while the other term increases.
These behavior implies an extremum condition for $E_{surf}$ versus $a$.
We take this extremum to be at $a = 0.53$fm and use this to determine $B$.
This extremum condition on $E_{surf}$ leads to the result that
the gradient term and the the reduction in binding energy from
diffuseness make equal contributions to $E_{surf}$ at its saturation.
For $a=0.53$fm, with the approximated form of $E_{grad}$,
the $B = 85.47$MeV$\cdot$fm$^5$ and $E_{diff} = E_{grad} = 10.36 A^{2/3}$ MeV.
The surface energy $E_{surf}$ is the same form as Eq.(\ref{esuf})
with modified coefficient of $a_0$ term to be 542.43, i.e.,
\begin{eqnarray}
   E_{surf}(R)/A &=& - (a/R) (65.18A^{2/3}/R^2-542.43A/R^3+459.65A^{4/3}/R^4)  
\end{eqnarray}
The incompressibility will also have two contributions each
from $E_{diff}$ and $E_{grad}$.
To have the same surface energy $E_{surf} = E_{diff} + E_{grad}$
for other values of $\alpha$, $a$ can be determined to have
 $E_{diff} = -(4\pi\rho_0/3)^{1/3} (a/A^{1/3}) (21\times 2.2769 
           - (37+23/\alpha) 3 + (23/\alpha) C_{\alpha 1} ) = 10.36/A^{1/3}$
with $C_{\alpha 1}$ the coefficient of $a/R$ term in $C_\alpha$.
Then the parameter $B$ can be set by
 $E_{grad} = \frac{3 B}{8\pi a^2} \left(\frac{4\pi\rho_0}{3}\right)^{4/3}
 \frac{a}{A^{1/3}} = 10.36/A^{1/3}$.

The values of $\kappa$ at the minimum $E/A$ can be obtained 
from Eq.(\ref{eras}) with $E_{grad}$ included.
The energy $E(R)/A$ with $E_{grad}$ is
\begin{eqnarray}
 E(R)/A &=& 21 C_K(a) (R_0/R)^2/D^{5/3} - (37+23/\alpha) C_0(a) (R_0/R)^3/D^2
      + (23/\alpha) C_\alpha(a) (R_0/R)^{3\alpha+3}/D^{\alpha+2}
            \nonumber \\  & &
      + 10.36 C_G(a) (R_0/R)^4/D^2
\end{eqnarray}
where
\begin{eqnarray}
 C_K(a) &=& 1 - 2.2769 (a/R) + 9.1046 (a/R)^2 - 7.805 (a/R)^3    \nonumber \\
 C_0(a) &=& 1 - 3 (a/R) + \pi^2 (a/R)^2 - \pi^2 (a/R)^3           \nonumber \\
 C_{1/3}(a) &=& 1 - 3.585 (a/R) + 10.819 (a/R)^2 - 11.642 (a/R)^3      \\
 C_1(a) &=& 1 - (9/2) (a/R) + (\pi^2+3) (a/R)^2 - (3\pi^2/2) (a/R)^3  
                      \nonumber  \\
 C_G(a) &=& 1 + (\pi^2/3-2) (a/R)^2
                      \nonumber  \\
 D(a) &=& 1 + \pi^2 (a/R)^2         \nonumber
\end{eqnarray}
Since $E_{surf}$ is not a small quantity we need exact calcuration here.
For $A=216$, 125 and 64 with $a = 0.53$fm, the respective values of $\kappa$ 
are 145, 128 and 103 in MeV, at the minimum values $R_{m}$ of $E/A$ 
 $R_{m}=6.888$, 5.709 and 4.527 or $x_m=1.0518$, 1.0691 and 1.0981 respectively.
These values are well below the value of $\kappa=234$ without finite
size corrections. An approximate expression for these $\kappa$ is given by
\begin{eqnarray}
  \kappa=\kappa(0) - 2.26 \kappa(0)/A^{1/3}    \label{kapasurf}
\end{eqnarray}
while approximate calculation up to first order 
of $a/R = (4\pi\rho_0/3)^{1/3} (R_0/R) (a/A^{1/3})$ 
gives $\kappa = \kappa(0)(1 - 2.01/A^{1/3})$.
The form of this result not only applies for $\alpha=1/3$, 
but also for softer EOS or 
smaller $\alpha$, and harder EOS or larger $\alpha$. 
The $\kappa(0)$ is $\kappa$ at $R\to\infty$ which is the infinite $A$ $\kappa$.
Thus $\kappa(0)=165$, 234, 372 for $\alpha=0$, 1/3, 1 respectively.
The simple result given by the equation comes from the observation that 
the surface correction coefficient 
scales with the $\kappa(0)$ coefficient for different $\alpha$.
This observation came from the detailed calculation that were done 
for various $A$ from $A=64$ to $A=216$ and various
values of $\alpha$. For example, when $\alpha=1$ $\kappa(0)=372$
and detailed calculations gave $\kappa=249$, 224, 188  
for $A= 216$, 125, 64 with $a=0.478$fm which can be approximated as 
 $\kappa=\kappa(0)(1 - 1.98/A^{1/3})$.   

The temperature and entropy dependence of the incompressibility can be 
evaluated using Eq.(\ref{kappaex}) for the $T$ dependent kinetic energy
with the $\kappa_0$ and $Sk$ determined above for the surface energy included.
Beside the kinetic energy, the surface energy itself can have
a temperature dependence which needs further investigation.

\subsection{
Symmetry Energy Contribution}

    We next investigate the role of symmetry energy terms. 
Symmetry energy terms appear in both the kinetic energy and potential energy
parts of the total energy. The symmetry energy appears as $(N-Z)^2/A^2$ in 
the binding energy per particle in the well known mass formulae. 
At low temperatures, $T \ll E_F$, and for proton fractions $y=Z/A$, 
such that $|2y-1| \ll 1$, the kinetic energy per particle is
\begin{eqnarray}
 E_K/A = 21x^{2/3} (1+(5/9)(2y-1)^2)+(\pi^2/140) 
                   (T^2/x^{2/3}) (1-(1/9) (2y-1)^2). 
           \label{kpaylt}
\end{eqnarray}
At higher temperatures, an expansion in $\rho \lambda^3$ is more appropriate 
and the kinetic energy per particle now becomes \cite{sjllg}
\begin{eqnarray}
 E_K/A=(3T/2) \{1+c_1 (\lambda^3 \rho_0/4) x (1+(2y-1)^2)    
           + c_2 (\lambda^3 \rho_0/4)^2 x^2 (1+(2y-1)^2)+\cdots\}.
                  \label{kpayht}
\end{eqnarray} 
The $y$ dependence in Eq.(\ref{kpayht}),
to the expansion order $c_2$, is not restricted to $|2y-1| \ll 1$,
i.e, it is valid for any $y$.  The potential energy per particle, $U/A$, 
is \cite{sjllg}
\begin{eqnarray}
    U/A = -b_0 x + b_\alpha x^{1+\alpha} + (2/3) (1/2+x_0) b_0 (2y-1)^2 x  
              -(2/3) (1/2+x_3) b_\alpha (2y-1)^2 x^{\alpha+1}    \label{ppay}
\end{eqnarray}
The $x_0$ is a parameter which we fix to give the empirical symmetry 
energy coefficient in the binding energy mass formula.  
This symmetry energy coefficient is taken to be 25 MeV. Note that the kinetic
energy contributes to this value and from Eq.(\ref{kpaylt}) is (35/3) MeV 
at $x=1$ and $T=0$.
The remaining  (40/3) MeV comes from the third and fourth term  
on the rhs of Eq.(\ref{ppay}).
Thus $(40/3)= (2/3) (1/2+x_0) b_0 - (2/3) (1/2+x_3) b_\alpha$ 
then $x_0$ and $x_3$ satisfy $(37+23/\alpha) x_0 - (23/\alpha) x_3 = 3/2$.
For $\alpha=1/3$, $106 x_0 - 69 x_3 = 3/2$.
If $x_3 = 1$, which is the value taken for Skyrme 
forces ZR1, ZR2, ZR3, then $x_0=0.6651$.
If $x_3=-1/2$, so that the fourth term in Eq.(\ref{ppay}) is absent, 
then $x_0=-0.3113$.
If $x_0=-1/2$, so that the third term is now absent, then $x_3=-0.7899$.
The nuclear incompressibility involves the second derivative of $E/A$ 
with respect to $x$ or $\rho$ so that the fourth term in Eq.(\ref{ppay}), 
which involves $x^{\alpha+1}$, is very important in the asymmetric part of
the nuclear incompressibility.  
Terms involving linear powers in $x$ do not make explicit contributions, 
but they do affect the minimum point in the  evaluation of $E/A$. 
However because of the uncertainty in the value of $x_3$, and thus in the
coefficient $1/2+x_3$ that appears in the fourth term, the contribution of 
the asymmetry term to the nuclear incompressibility would seem to be 
somewhat uncertain. 
This turns out not to be the case because of cancellations so that the 
coefficient multiplying the $(1/2+x_3)$ part in $\kappa$ is small 
as will be found.
We can evaluate the relative  importance of the contribution of the symmetry 
energy to $\kappa$ by isolating those terms associated with it 
and turning off the Coulomb and surface contributions. 
Let us consider the choice of a general $x_3$, which has 
an associated $x_0$ to give correct symmetry energy. 
Our $E(R)$ with symmetry terms included and is
\begin{eqnarray}
   E(R)/A = C_K(y) (R_0/R)^2 - C_0(y) (R_0/R)^3
                + C_\alpha(y) (R_0/R)^{3(\alpha+1)}
\end{eqnarray}
with                         
\begin{eqnarray}
  C_K(y) &=& 21 + (35/3) (2y-1)^2                     \nonumber \\
  C_0(y) &=& (37+23/\alpha) - (40/3 + (46/3\alpha) (1/2+x_3)) (2y-1)^2
              \nonumber \\  
  C_\alpha(y) &=& (23/\alpha) - (46/3\alpha) (1/2+x_3) (2y-1)^2       
\end{eqnarray}
Thus 
\begin{eqnarray}
  0 &=& - 2 C_K(y) (R_0/R)^2 + 3 C_0(y) (R_0/R)^3
        - (3\alpha+3) C_\alpha(y) (R_0/R)^{3\alpha+3}    \\
  \kappa &=& 6 C_K(y) (R_0/R)^2 - 12 C_0(y) (R_0/R)^3   
           + (3\alpha+3) (3\alpha+4) C_\alpha(y) (R_0/R)^{3\alpha+3}
                    \label{kappayr}   \\
  Sk &=& - 24 C_K(y) (R_0/R)^2 + 60 C_0(y) (R_0/R)^3   
         - (3\alpha+3) (3\alpha+4) (3\alpha+5) C_\alpha(y) (R_0/R)^{3\alpha+3}
\end{eqnarray}
For $\alpha=1/3$,
the minimum point of $E(R)/A$ is determined by a simple quadratic equation.
The value of $\kappa$ at this minimum is accurately given by, 
up to $(2y-1)^2$ order,
\begin{eqnarray}
   \kappa = 234 - (300 - 16 (1/2+x_3)) (2y-1)^2  
\end{eqnarray}
with $x_m = (R_0/R_m)^3 = (1 - (0.271 - 0.197(1/2+x_3))(2y-1)^2)^3$.
The small coefficient of 16 compared to 300 in front of the term 
involving $x_3$, means that uncertainties in $x_3$ do not have a 
pronounced effect on the value of $\kappa$.
Similar results are also true for other values of $\alpha$,
with the second $1/2+ x_3$ term about 5\% of the first term 
in the $(2y-1)^2$ dependent part of $\kappa$.
For general $\alpha$, using either Eq.(\ref{kappayr}) or (\ref{kappaex}),
\begin{eqnarray}
 \kappa &=& \kappa_0(\alpha)
   - \frac{182.9 + 403.33\alpha + 190\alpha^2}{0.7971+\alpha} (2y-1)^2
   + \frac{(1/2+x_3)(9.333+28\alpha)}{0.7971+\alpha} (2y-1)^2
        \label{kappay}
\end{eqnarray}
with $x_m = (1 - (0.3060 - 0.2222(1/2+x_3))(2y-1)^2/(0.7971+\alpha))^3$.
Surface symmetry corrections can also be easily included following 
the procedure given above. 
Since the volume and volume symmetry terms have the same density dependence, 
the surface corrections to each will scale in exactly the same way with their
corresponding volume term. Thus we can rewrite the symmetry term in the above
equation as $(300-16 (1/2+x_3)) (2y-1)^2 (1-2.26)/A^{1/3})$ 
for $\alpha=1/3$,
where the $2.26/A^{1/3}$ part of this result is now the surface 
symmetry correction to the symmetry energy (c.f., Eq.(\ref{kapasurf})).  

At non zero but low $T$, Eq.(\ref{kpaylt}) gives a small $T$ dependent 
kinetic energy term $E_x = (\pi^2/140) T^2 (R/R_0)^2 (1 - (2y-1)^2/9)$.
Using Eq.(\ref{kappaex}),
\begin{eqnarray}
 \kappa(T) = \kappa(0) - (\pi^2/70) T^2 (1 - (2y-1)^2/9) (Sk/\kappa(0) + 1)
\end{eqnarray}
where $\kappa(0)$ is the incompressibility at zero $T$ for an asymmetric
system which is given by Eq.(\ref{kappay}).
On the other hand $E_x = (35/\pi^2) (S/A)^2 (R_0/R)^2 (1 - (2y-1)^2/9)$ 
for a given entropy since $S/A = (\pi^2/2) T/E_F$.
This gives the adiabatic incompressibility of
\begin{eqnarray}
 \kappa_Q(T) = \kappa(0) 
        + (70/\pi^2) (S/A)^2 (1 - (2y-1)^2/9) (Sk/\kappa(0) + 5)
\end{eqnarray}

At higher $T$, the energy is the sum of Eq.(\ref{kpayht}) and Eq.(\ref{ppay})
and we can neglect higher order terms in virial expansion keeping terms only 
up to $c_2$.
Then $\kappa$ is given by:
\begin{eqnarray}
 \kappa(T) = 27 \frac{c_2}{T^2}  \left(\frac{2\pi\hbar^2}{m}\right)^{3}
                \left(\frac{\rho_0}{4}\right)^2 x_m^2
                (1 + (2y-1)^2)
     + 9 \alpha (1+\alpha) b_\alpha
         \left[1 - \frac{2}{3}\left(\frac{1}{2}+x_3\right)(2y-1)^2\right]
            x_m^{1+\alpha}  \label{kappayt} 
\end{eqnarray}
The $x_m$ is again the minimum of  $E/A$, but now evaluated with the new
kinetic energy. The $x_m$ is affected by both the $c_1$ and $c_2$ terms.
A limiting value of $\kappa$ can be obtained by taking $T$ very large where
an ideal gas $E_K/A=(3/2)T$ leads to very simple results. 
Adding this $E_K/A$ to $U/A$
given by Eq.(\ref{ppay}) leads to a minimum $x_m$ given by
 $x_m^\alpha=b_0[1-(2/3)(1/2+x_0)(2y-1)^2]/
     (b_\alpha (1+\alpha) [1- (2/3)(1/2+x_3)(2y-1)^2])$.
In this high $T$ limit $\kappa$ is given by the second term on
the right side of Eq.(\ref{kappayt}) and is
\begin{eqnarray}
  \kappa(T\buildrel{>}\over{\sim} 10 {\rm MeV})
       = 9 \alpha b_0 \left[1 - \frac{2}{3}(\frac{1}{2}+x_0)(2y-1)^2\right]
          \left(\frac{b_0\left[1 - \frac{2}{3}(\frac{1}{2}-x_0)(2y-1)^2\right]}
          {(1+\alpha) b_\alpha
              \left[1 - \frac{2}{3}(\frac{1}{2}+x_3)(2y-1)^2\right]}
           \right)^{1/\alpha}.
\end{eqnarray}
At fixed entropy 
 $E_K/A= (3/2) C_S \rho^{2/3} (1 + \sum{c_n ([2\pi\hbar^2/m C_S]^{3/2} 
    (1 + (2y-1)^2)})$.
The resulting $E(R)/A$ has a structure similar to the result for a degenerate 
Fermi gas since both have a $\rho^{2/3}$ dependence for the kinetic energy
term but with different coefficients.
This feature and a similar result at lower $T$ suggests a duality 
in the energy per particle EOS at constant entropy and its associated 
$\kappa_Q$ and the $T=0$ EOS and its associated constant $T$ $\kappa$ 
even for asymmetric system.

\subsection{
Coulomb Energy Contribution}

  Next, let us consider the Coulomb energy contribution. 
For a uniformly charged sphere we have 
 $E_c/A = (3/5)e^2(Z^2/A)/R=(3/5)(e^2/\hbar c) (\hbar c)(Z^2/A)/R$.
As before, we turn off the other two contributions, now the surface and
symmetry, to first study the isolated importance of the Coulomb interaction. 
In this case our energy per particle at $T=0$ is  
\begin{eqnarray}
 E(R)/A = 28.627A^{2/3}/R^2 - (58.89+36.61/\alpha)A/R^3
        + (23/\alpha)(A/(0.8565R)^3)^{\alpha+1} + 0.8628(Z^2/A)/R 
\end{eqnarray}
The Coulomb contribution to $\kappa$ can be expressed in terms of
a simple result, using Eq.(\ref{kappaex}),
\begin{eqnarray}
    \kappa_C = 4 E_C /A + (E_C/A) (Sk/\kappa_0)
\end{eqnarray}
The skweness $Sk$, third derivative of $E_0(R)$, is negative. 
Here $E_0(R)$ is the energy without the Coulomb correction and $\kappa_0$  
its associated incompressibility. 
The rhs is evaluated at the minimum without
Coulomb corrections. For a uniform sphere of radius $R$, 
for $\alpha=1/3$, the  
\begin{eqnarray}
  \kappa = 234 - 4.75 Z^2/A^{4/3} 
\end{eqnarray}
For $A=216$, $Z=86$, $R_{m}=7.14$ and $\kappa=207$MeV; for $A=125$, $Z=50$,
$R_{m}=5.914$ and $\kappa=215$MeV;
for $A=64$, Z=28, $R_{m}=4.717$ and $\kappa=220$MeV. 
If the density profile is taken to that of a Wood-Saxon form, the Coulomb 
energy can be approximated by $(3/5)(Z^2/A)(1-(7 \pi^2/6) (a/R)^2)/R$
\cite{bohrmot}.
Thus surface contributions appear both directly and indirectly, 
modifying the other terms.
Adding surface contributions gives, using Eq.(\ref{kappaex}) for $\alpha=1/3$,
\begin{eqnarray}
     \kappa= 234 - 4.75Z^2/A^{4/3} + 22.9Z^2/A^2 
\end{eqnarray}
Other values of $\alpha$ can also be easily evaluated. 
Neglecting surface-Coulomb we have
for $\alpha=1$, $\kappa=372-6.25Z^2/A^{4/3}$ and
for $\alpha=0$, $\kappa=165-3.875Z^2/A^{4/3}$.
The ratio of the Coulomb coefficient of $Z^2/A^{4/3}$ to volume coefficient is 
not the same for each $\alpha$ so that a simple scaling result is not possible. 

Since both the Coulomb energy and the $T$ dependent part of kinetic enrgy 
are small the effects of these energies to the incompressibility are
additive.

\section{
Conclusions}

In this paper we investigated the behavior of the 
asymmetric nuclear matter 
incompressibility at finite temperature and entropy uing
a mean field theory.
The incompressibility is related to the curvature of the energy
as a function of density, with the isothermal incompressibility
having temperature held constant and the adiabatic incompressibility
has entropy held constant.
These two incompressibilities are the same at $T=0$ or $S=0$ 
but have very different behaviors with temperature $T$ and entropy $S$.
At all $T$ or $S$ these incompressibilities are
very sensitive to the choice of $\alpha$, the density dependent power of 
the repulsive interaction. 
We look at extreme cases of a very soft equation of state where
$\alpha$ goes to zero to a very hard equation of state with $\alpha = 1$.  
The range of $\kappa$ is from  165MeV  to  372MeV.  
For a moderate equation of state that is
frequently used, $\alpha = 1/3$ and $\kappa= 234$MeV.  
When a temperature dependence is included, the value of the isothermal 
incompressibility $\kappa$ increases sharply with $T$.
At low $T$ or $T$ much less than the Fermi energy, $\kappa$ increases as $T^2$.
For example at $T= 7.5$MeV and $\alpha =1/3$, $\kappa$ is 302MeV.
By contrast, the adiabatic incompressibility was shown to decrease with 
increasing entropy and it eventually goes to zero.
The adiabatic incompressibility is shown to arise from 
an equation of state (EOS) or energy per particle that has a
structure that is similar to a $T=0$ Fermi gas.

The incompressibility is also studied in finite systems 
where we investigated the role of surface, symmetry energy
and Coulomb energy effects. First we studied each of these terms separately 
to illustrate the relative importance of them. 
Of these three terms, the surface had the largest effect, 
followed by the Coulomb and then the symmetry term for nuclei
in the valley of $\beta$ stability. 
Uncertainties in the $x_3$ parameter of the Skryme interaction were 
shown not to be so important. In our study of the surface term
we used a Wood-Saxon density distribution. We found an approximation scheme 
to do various integrals (similar to the Sommerfeld method for Fermi gas 
integrals) that appear in our analysis and obtained very accurate analytic 
results for the surface contribution to the energy per nucleon.

This work was supported 
in part by the DOE Grant No. DE-FG02-96ER-40987 and DE-FG02-95ER-40940 and
in part by Grant No. R05-2001-000-00097-0 from the Basic Research Program 
of the Korea Science and Engineering Foundation.

\end{document}